# Black Start Operation of Grid-Forming Converters Based on Generalized Three-phase Droop Control Under Unbalanced Conditions

Zexian ZENG*
Georgia Institute of Technology
USA
zzeng88@gatech.edu

Prajwal BHAGWAT
University of Wisconsin-Madison
USA
pbhagwat2@wisc.edu

Maryam SAEEDIFARD
Georgia Institute of Technology
USA
maryam@ece.gatech.edu

Dominic GROSS
University of Wisconsin-Madison
USA
dominic.gross@wisc.edu

**SUMMARY**

Grid-forming (GFM) converters are envisioned as the cornerstone of the future power systems, driven by the increasing integration level of inverter-based resources. Unlike grid-following (GFL) converters, which depend on a stable ac voltage waveform for synchronization, GFM converters impose a stable ac voltage waveform at their terminals and self-synchronize through the grid, emerging as a potential substitute for synchronous generators in the future power system. To accomplish this, GFM converters must provide ancillary services typically offered by synchronous generators, including the ability to black start a system. This paper focuses on the challenging task of bottom-up restoration in a complete blackout system using GFM converters. Challenges arise due to the limited current capability of power converters, resulting in distinct dynamic responses and fault current characteristics compared to synchronous generators. Additionally, GFM control needs to address the presence of unbalanced conditions commonly found in distribution systems. To address these challenges, this paper explores the black start capability of GFM converters with a generalized three-phase GFM droop control. This approach integrates GFM controls individually for each phase, incorporating phase-balancing feedback and enabling current limiting for each phase during unbalanced faults or overloading. The introduction of a phase-balancing gain provides flexibility to trade-off between voltage and power imbalances. The study further investigates bottom-up black start operations using GFM converters, incorporating advanced load relays into breakers for gradual load energization without central coordination. The effectiveness of bottom-up black start operations with GFM converters, utilizing the generalized three-phase GFM droop, is evaluated through electromagnetic transient (EMT) simulations in MATLAB/Simulink. The results confirm the performance and effectiveness of this approach in achieving successful black start operations under unbalanced conditions.

# 1. Introduction

Due to the growing global concerns over climate change and the ever-increasing demand for electricity, the integration of substantial renewable energy into power systems has attracted significant traction. This transition underscores the pivotal role of power electronics in interfacing renewables and the grid, establishing a foundation for resilient and sustainable power systems in the future [1]. Advanced power converter control has the potential to replace various functions traditionally associated with conventional synchronous generators (SGs). In future power systems characterized by a high integration level of inverter-based resources (IBRs), a key focal point is understanding how these power converters can effectively contribute to providing black start support.

Control strategies for grid-connected power converters fall into two categories [1], [2]. Firstly, there are grid-following (GFL) strategies, which may offer grid-supporting services but require the presence of other devices to establish a stable ac voltage waveform for synchronization. Secondly, there are grid-forming (GFM) control strategies that autonomously impose a well-defined and stable ac voltage waveform at their point of interconnection, self-synchronizing through the grid. These GFM strategies are widely envisioned as the cornerstone of future grids characterized by a high integration level of IBRs. Given their ability to establish a stable ac voltage waveform at their terminals, GFM converters can provide black start support, effectively supplanting SGs.

Recent developments highlight the bottom-up restoration of a complete blackout system without energization from the upstream network, by leveraging GFM converters [3]-[5]. Explorations into advanced protection and black start mechanisms tailored to the black start capabilities of GFM have recently been investigated in [6] without traditional communication or overlaid control. However, implementing black start operation in a power system using GFM converters introduces technical challenges rooted in the inherent differences between power converters and traditional SGs. The first challenge arises from the limited current capability of power converters. Replicating the overcurrent response exhibited by SGs becomes challenging for GFM converters [7], particularly when faced with startup currents exceeding device-rated values, as commonly observed in induction motors. Additionally, in the context of distribution system black starts, addressing system imbalance caused by uneven loading is crucial. The imbalance of distribution feeders may further aggregate during the sequential recovery stage of a black start. The majority of existing studies on black start operations with GFM converters often neglect the substantial imbalances prevalent in distribution systems. Studies addressing GFM control under unbalanced conditions and incorporating current limiting often rely on symmetrical components. However, limiting phase current through the control of symmetrical components introduces a highly nonlinear relationship, posing challenges in control design [8]. For example, the method in [9] leverages symmetrical components with current limiters to limit the currents in the *abc* frame. Nonetheless, this method requires the calculation of the Root Mean Square (RMS) value for each phase current, a procedure that introduces delays and may result in significant overcurrent occurrence for durations exceeding one cycle.

To address these issues, this paper studies GFM converters using a generalized three-phase GFM droop control, as proposed in [8], for black starting an unbalanced system. This control scheme integrates individual GFM controls for each phase with phase-balancing feedback, enabling current limiting for unbalanced faults or overloading. The additional control degree introduced by the phase-balancing gain provides flexibility to trade-off between voltage and power imbalance under unbalanced conditions, effectively utilizing the inherent degrees of freedom in power converters. Moreover, this paper explores bottom-up black start scenarios with advanced load relays integrated into breakers, following the methodology presented in [6]. This approach facilitates a gradual energization of loads without requiring central coordination.

To illustrate the effectiveness of the black start study using GFM converters with the generalized three-phase GFM droop control, this paper conducts electromagnetic transient (EMT) simulations in MATLAB/Simulink. The IEEE 13 bus feeder system is utilized as the benchmarking system, incorporating advanced load relays integrated into breakers. Additionally, an induction motor model is integrated to capture inrush currents. The results demonstrate successful black start operation using GFM converters with the generalized three-phase GFM droop control. The phase-balancing gain enables a trade-off between voltage and power imbalance, while ensuring accurate active power sharing among different GFM converters for efficient system restoration. Furthermore, the implemented current limiter proves effective in limiting the current of each phase under unbalanced conditions.

## 2. Generalized Three-phase GFM Droop Control and EMT Models for Black Start Study

This section begins with an overview of the generalized three-phase GFM droop control, incorporating current limiters for each phase. Subsequently, EMT models for loads and induction motors are introduced for the black start study.

### 2.1. Inverter Models Based on Generalized Three-phase GFM Droop Control

The GFM converter, as depicted in Fig. 1, is a two-level three-phase dc/ac voltage source converter (VSC) with a grounded dc-side midpoint. For the sake of brevity, the dc-side dynamics are neglected in this paper. The terminal voltage, filter voltage and current, and output current are represented by $v_{\text{sw}}$, $v$, $i$, and $i_o$, respectively. The control architecture of the GFM-VSC is illustrated in Fig. 1. To address unbalanced conditions and black start scenarios, this paper employs the generalized three-phase GFM control. As detailed in [8] and depicted in Fig. 1, this control strategy integrates individual per-phase inner current and voltage controllers with GFM droop control and includes phase-balancing feedback.

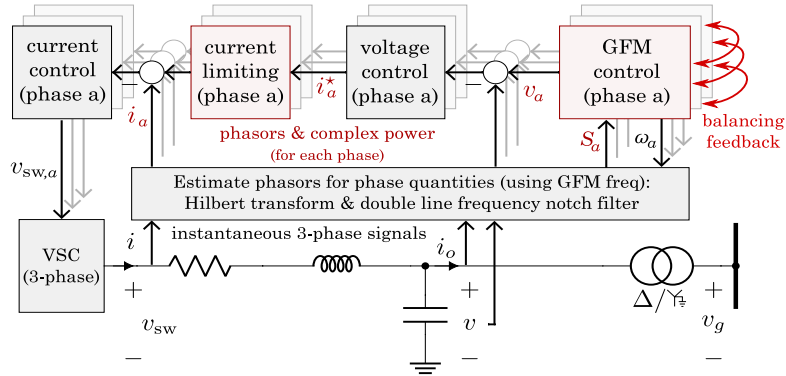

Fig. 1. The generalized three-phase GFM control architecture [8].

#### a. Generalized Three-phase GFM Droop Control

The generalized GFM droop control integrates single-phase droop control for each phase $p \in \mathcal{P}$, where $\mathcal{P} \coloneqq \{a, b, c\}$, with phase-balancing feedback. Considering $\left(\theta_a^{\text{bal}}, \theta_b^{\text{bal}}, \theta_c^{\text{bal}}\right) \coloneqq \left(0, -\frac{2}{3}\pi, \frac{2}{3}\pi\right)$ and voltage setpoints $V_p^\star$, the GFM voltage angle and magnitude references $\theta_p^{\text{gfm}} = \delta_p^{\text{gfm}} + \theta_p^{\text{bal}}$ and $V_p^{\text{gfm}}$ are determined by

$$\frac{\mathrm{d}}{\mathrm{d}t}\delta_p^{\text{gfm}} = \omega_0 - \sum_{l \in \mathcal{P} \setminus p} k_s \left(\delta_p^{\text{gfm}} - \delta_l^{\text{gfm}}\right) + m_P(P_p^\star - P_p), \tag{1a}$$

$$V_p^{\text{gfm}} = V_p^\star - \sum_{l \in \mathcal{P} \setminus p} k_s \left(V_p^{\text{gfm}} - V_l^{\text{gfm}}\right) + m_Q(Q_p^\star - Q_p), \tag{1b}$$

where $P_p^\star$ and $Q_p^\star$ are the active and reactive power setpoints, and $\omega_0$ is the nominal frequency. The active and reactive power measurements for each phase are denoted by $P_p$ and $Q_p$. Moreover, $\omega_p^{\text{gfm}} =$ $\omega_0 + \frac{d}{dt}\delta_p^{\text{gfm}}$ is the GFM reference frequency for each phase $p \in \mathcal{P}$, $m_P$ is the $P - f$ droop coefficient, $m_Q$ is the $Q - V$ droop coefficient, and $k_s$ is the phase-balancing gain.

This paper delves into the utilization of phase-balancing feedback to improve black start operation, offering a trade-off between voltage and power imbalance under unbalanced conditions. The adjustable parameter $k_s$ enables the reduction of power imbalance at smaller $k_s$ values, while promoting more balanced voltage at larger $k_s$ values [8]. Furthermore, the generalized three-phase GFM droop control seamlessly converges to the standard droop control as $k_s$ approaches infinity. The introduced degree of freedom via $k_s$ provides flexibility in achieving a trade-off between voltage and power imbalance during the black start of an unbalanced system.

**b. Dual-loop Current/Voltage Control and Current Limiting**

As illustrated in Fig. 1, the standard dual-loop proportional-integral (PI) current and voltage control, incorporating reference current limiting, is extended to each phase to track the GFM voltage references. To facilitate this process, the Hilbert transform is implemented to generate the quadrature components of voltage and current for each phase. Subsequently, the voltage and current of each phase are transformed to a rotating frame attached to the angle references generated by the generalized three-phase GFM droop control [8].

During unbalanced overloading and fault scenarios, relying on the standard current-limiting method, which limits the three-phase average current, may lead to suboptimal performance [8], potentially causing the inverter to exceed the current rating of semiconductors. To overcome this challenge, the current-limiting method is individually extended for each phase. In this approach, the current limiter adjusts the magnitude of the sinusoidal reference current for each phase. It actively limits the current magnitude of each phase, ensuring that asymmetrical faults or overloading do not compromise the performance of healthy phases.

## 2.2. Load Models

The benchmark system chosen for this study is the IEEE 13 bus feeder, serving as a representation of an unbalanced distribution system. Unbalanced loads are portrayed using constant impedance load models. Furthermore, the distribution system incorporates balanced three-phase dynamic loads. When terminal voltages fall below 0.7 p.u., these loads are characterized by a constant impedance representation. Conversely, for voltages exceeding 0.7 p.u., the loads are modeled as constant current loads.

## 2.3. Induction Motor Model

The energization/initiation of induction motor loads poses distinct challenges compared to static loads. To investigate the response of GFM-VSCs during the startup of induction motor loads with limited current capability, this study uses a dynamic model of a three-phase induction motor.

## 3. Bottom-up Black Start Under Unbalanced Conditions

This section outlines a strategy for bottom-up black start, eliminating the requirement for centralized coordination. It introduces an advanced load relay and describes a black start methodology using GFM-VSCs for unbalanced conditions. This methodology can serve as a guideline for future power system black start operations.

## 3.1. Breakers with Advanced Load Relays

To enhance black start capabilities, this study investigates the integration of advanced load relays within breakers. These relays combine the functionalities of frequency and voltage relay, tripping or resetting the breaker based on predefined frequency or voltage operating ranges after specified delay times, as illustrated in [6]. The advanced load relays, integrated into the breakers, are designed for load connection/disconnection after specific connection/disconnection waiting times $T_{wait,con}$/$T_{wait,dis}$ during the black start operation.

Frequency serves as an indicator of available power, aligning with the droop behavior of the GFM-VSCs. When the frequency surpasses $f_{min}$, the available power is deemed sufficient for load connection, and conversely, when it falls below $f_{min}$, loads must be disconnected. Simultaneously, the voltage relay monitors the voltage magnitude continuously, initiating load disconnection if the terminal voltage magnitude exceeds $V_{max}$ or falls below $V_{min}$.

By utilizing individual reset delay times for load relays, the system operator can achieve a gradual energization/connection of loads during the black start process, thereby mitigating the risk of Cold Load Pickup [6] associated with inrush currents. These load relays continuously monitor the frequency and voltage as indicators of the grid's state, thereby eliminating the need for additional communication with the inverters during the black start process. The gradual energization, facilitated by the integration of advanced load relays into the breakers, is implemented in the black start process of this study. The waiting times of individual relays are initialized to prevent simultaneous switching of all loads on/off, thus mitigating inrush current. Furthermore, these waiting times are adjustable, offering system operators flexibility based on load priorities. It is important to note that sequential "bottom-up restoration" stage can be tailored by the system operator to align with various economic objectives while accommodating system operation constraints [3]. Various optimization problems have been proposed to optimize the overall system operating point and topology (see, e.g., [10]), a topic that falls beyond the scope of this work.

## 3.2. Definite-time Overcurrent Relays

Alongside the advanced load relays, definite-time overcurrent relays are integrated at the loads for protection purposes. In essence, if the current value exceeds the predetermined threshold for a specific duration, the overcurrent relay issues a trip signal to the breaker, thereby disconnecting the load from the system. In the simulations, five fundamental cycles are used to evaluate the fault current limiting capability of the GFM-VSCs.

## 3.3. GFM Black Start Logic

The GFM converter with the generalized three-phase GFM droop control emerges as a promising solution for black start operations in an unbalanced system. Following a blackout event, the leading GFM converter in the system gradually increases its terminal voltage by ramping up its voltage magnitude setpoints. This method achieves soft transformer energization and eliminates the inrush current. To prevent power oscillation among different GFM converters, a phase-locked loop (PLL) is used prior to connecting the second GFM converter [6]. The PLL initially synchronizes with the voltage at the connection point of the second converter. Following an appropriate synchronization period, the angle measured by the PLL is used to initialize the GFM control of the second GFM converter. With the synchronization and connection of all GFM converters in the system, the advanced relays are activated. As a result, the gradual reconnection of loads occurs during the black start operations based on local voltage and frequency information.

## 4. Simulation Results

To investigate the black start capability of GFM-VSCs with the generalized three-phase GFM droop control, EMT simulations are conducted using Simscape in the MATLAB/Simulink environment. The simulated system with its key parameters is shown in Fig. 2. Two GFM-VSCs are connected to the IEEE 13 bus feeder system with $\Delta-Y_g$ transformers (480 V $\Delta$ to 4.16 kV $Y_g$). Both converters have identical power ratings of 3 MW and droop coefficients of $m_P = 5\%$ and $m_Q = 5\%$. However, GFM-VSC 1 is configured with a phase-balancing gain $k_s$ of $10^5$, while GFM-VSC 2 operates with a phase-balancing gain $k_s$ of 1. Breakers equipped with relays including advanced load relays and definite-time overcurrent relays, as depicted in Fig. 2, are placed to automatically trip/reset loads based on the grid's state, ensuring a gradual connection of loads. Additionally, a 1 MVA three-phase induction motor is included in the system.

### 4.1. Black Start with GFM-VSCs

This case study investigates the restoration of the IEEE 13-bus feeder system following a total blackout. Simulation results, presented in Fig. 3, demonstrate the black start capability of GFM-VSCs with different $k_s$ values and the controlled gradual connection/energization of loads through advanced load relays. Using phase phasor extraction with the Hilbert transform [8], the magnitude of current and voltage as well as the power of each phase are calculated and plotted in Fig. 3.

The leading GFM-VSC 1 initiates a soft start at $t = 3$ s by gradually increasing its terminal voltage, ramping up its voltage magnitude setpoints. This approach ensures a soft transformer energization, effectively eliminating the inrush current. Prior to connection of GFM-VSC 2, GFM-VSC 1 independently supplies the base load. Upon the connection of GFM-VSC 2 at $t = 5$ s, the output power is evenly shared between the two GFM-VSCs. At $t = 8$ s, breakers with advanced load relays are activated, initiating a gradual reconnection of loads based on local voltage and frequency information. At $t = 21$ s, a zero-impedance line-to-ground fault occurs for phase $a$ at L692, as depicted in Fig. 2. The affected load is disconnected from the system by the overcurrent relay integrated into the breaker, the detailed waveforms of which will be presented in the subsequent section. The connection of the induction motor at $t = 27$ s introduces significant transients, whose detailed waveforms will also be presented in the following section.

Examining the active power for each phase of GFM-VSC 1 and GFM-VSC 3 in Fig. 3 reveals that GFM-VSC 1 with $k_s = 10^5$ exhibits imbalanced active power among its three phases. In contrast, GFM-VSC 2 with $k_s = 1$ shows a significantly improved balance in active power. Despite differences in active power levels among the three phases, the total active power is accurately shared between the two GFM-VSCs. This underscores the efficacy of the generalized three-phase GFM droop control in achieving active power sharing while adjusting power imbalances through tuning the phase-balancing gain during black start operations.

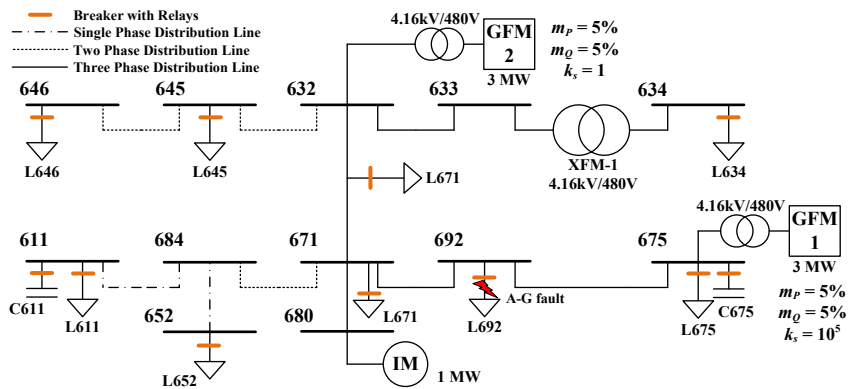

Fig. 2. The IEEE 13-bus feeder system.

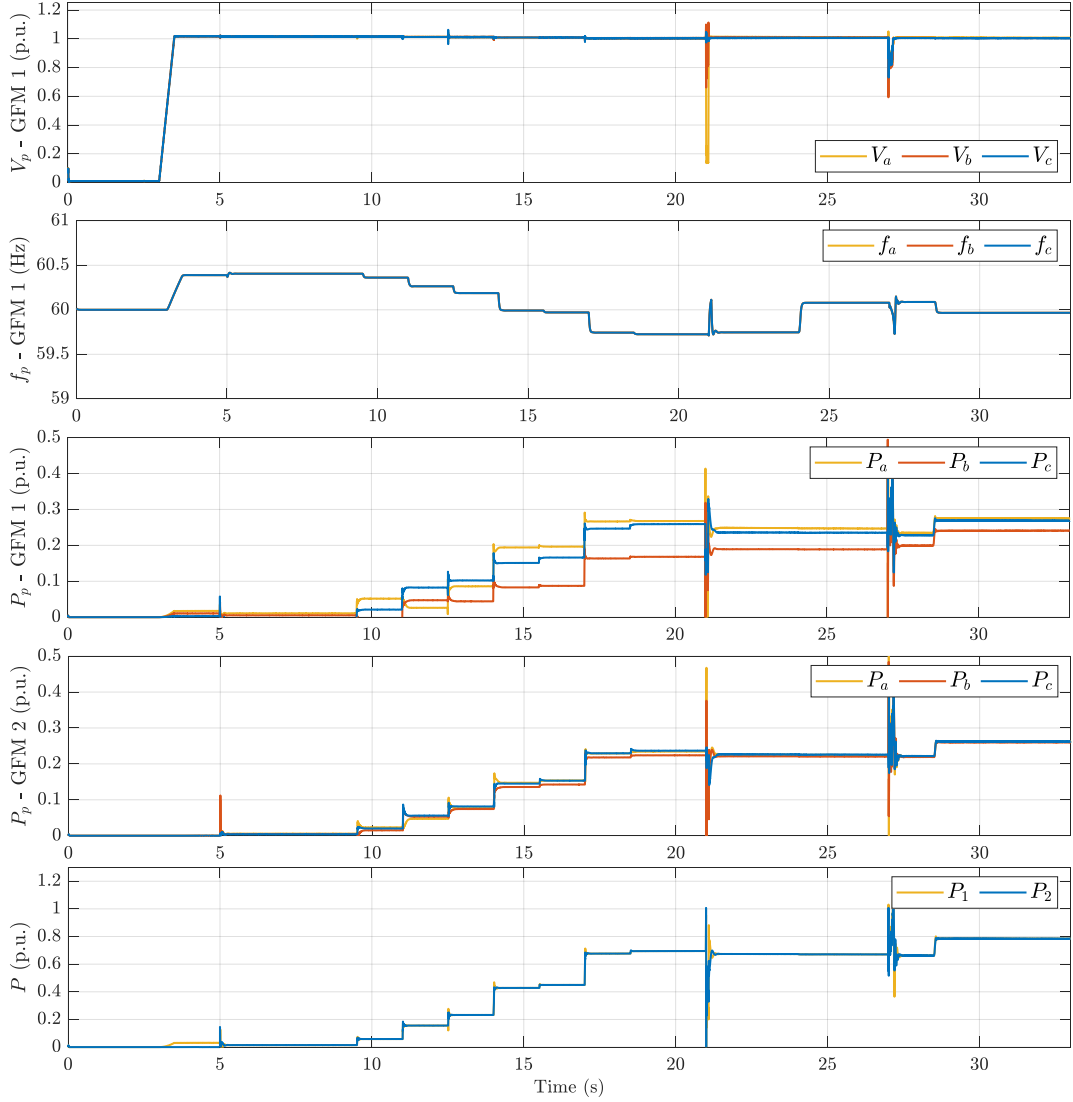

Fig. 3. Simulation results for the black start of IEEE 13-bus feeder system: voltage magnitude and frequency for each phase of GFM-VSC 1, active power for each phase of GFM-VSC 1 and GFM-VSC 2, and total active power of GFM-VSC 1 and GFM-VSC 2.

### 4.2. Impact of Phase-balancing Gain $k_s$

The phase-balancing gain $k_s$ introduces a degree of freedom, offering flexibility in achieving a trade-off between voltage and power imbalance during the black start of an unbalanced system. To illustrate the role of the phase-balancing gain $k_s$, the steady-state unbalance factors of GFM-VSC 2 versus its phase-balancing gain $k_s$ are presented in Fig. 4.

Let $V^+$ and $V^-$ denote the magnitudes of positive and negative sequence voltages, respectively, and $\bar{P} \coloneqq \frac{1}{3}\sum_{p\in\mathcal{P}} P_p$ and $\bar{Q} \coloneqq \frac{1}{3}\sum_{p\in\mathcal{P}} Q_p$ represent the average phase power of the converter, all in per unit. As defined in [8], the voltage unbalance factor is given by $V_{\text{UF}} \coloneqq V^-/V^+$. The power unbalance factors are defined as $P_{\text{UF}} \coloneqq \max_{p\in\mathcal{P}}|P_p - \bar{P}|$, and $Q_{\text{UF}} \coloneqq \max_{p\in\mathcal{P}}|Q_p - \bar{Q}|$. Figure 4 illustrates the steady-state unbalanced factors for the terminal voltages and power of GFM-VSC 2 in the IEEE 13-bus feeder system, with varying phase-balancing gain $k_s$. Meanwhile, GFM-VSC 1 maintains a constant phase-balancing gain of $10^5$. The results highlight a trade-off between voltage and power imbalances, i.e., reducing the phase-balancing gain $k_s$ decreases power unbalance factors but increases voltage unbalance factor.

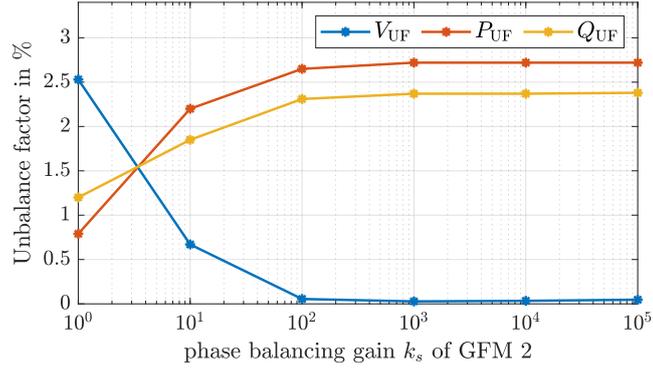

Fig. 4. Steady-state unbalance factors of GFM-VSC 2 versus its phase-balancing gain $k_s$.

### 4.3. Current Limiting During Single Line-to-ground Fault

A zero-impedance line-to-ground fault occurs for phase *a* at L692 at *t* = 21 s. The fault is rectified after five cycles by disconnecting L692 through the overcurrent relay. To demonstrate the capabilities of GFM-VSCs with the generalized three-phase GFM droop control in limiting the unbalanced fault currents, the responses of the filter current, filter voltage, active power, reactive power, and frequency for each phase of GFM-VSC 1 and GFM-VSC 2 are illustrated in Figs. 5 and 6, respectively. The current limiter for each phase successfully limits the current magnitude to $I_{max}$ = 1.2 p.u.

Notably, the frequency of the three phases is coherent for GFM-VSC 1 with $k_s = 10^5$, similar to the behavior of standard positive-sequence droop control. In contrast, for GFM-VSC 2 with $k_s = 1$, the frequency of the three phases diverges obviously, as depicted in Fig. 6. This divergence is attributed to the small value of the phase-balancing gain $k_s$. This behavior can be explained by (1a), i.e., a larger phase-balancing gain $k_s$ results in faster and stiffer synchronization of the angles $\delta_p^{\text{gfm}}$, where $\mathcal{P} \coloneqq \{a, b, c\}$, resulting in decreased voltage unbalance. Conversely, a small phase-balancing gain $k_s$ results in slower and less aggressive frequency/angle synchronization contributing to increased voltage unbalance but decreased power unbalance between the phases.

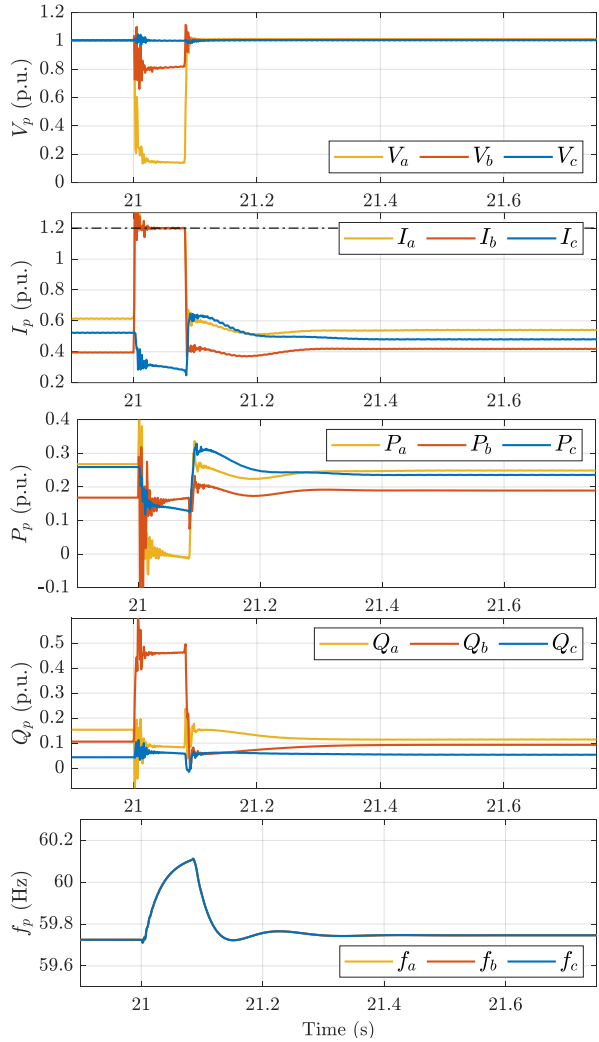 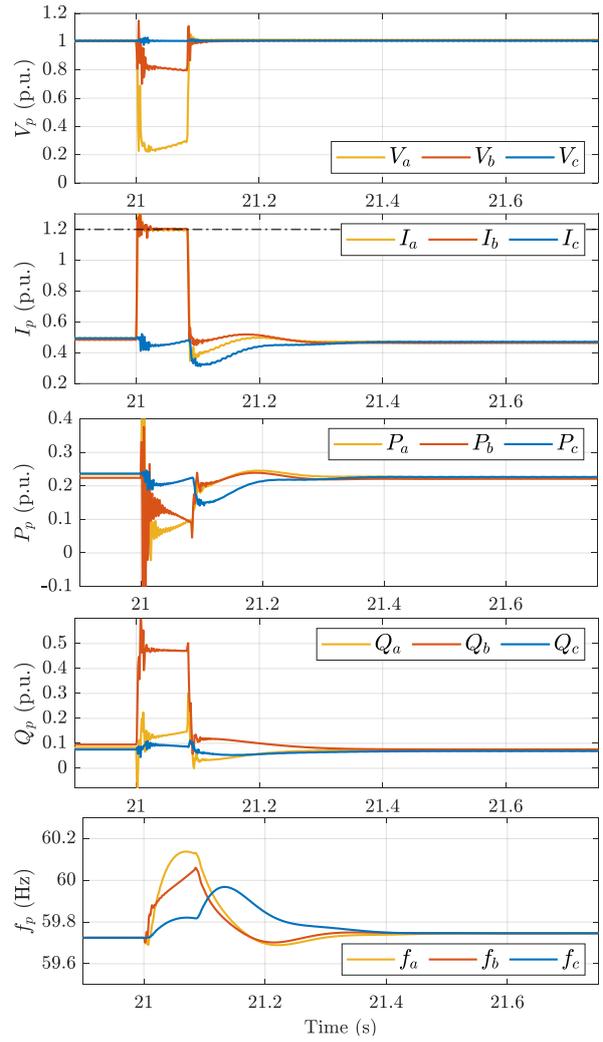

Fig. 5. The responses of the filter current, filter voltage, active power, reactive power, and frequency for each phase of GFM-VSC 1 to a phase *a* to ground fault on L692.

Fig. 6. The responses of the filter current, filter voltage, active power, reactive power, and frequency for each phase of GFM-VSC 2 to a phase *a* to ground fault on L692.

### 4.4. Current Limiting during Induction Motor Startup

The induction motor is connected to the system at $t = 27$ s, initially with zero load torque. Subsequently, at $t = 28.5$ s, a load torque of 0.6 p.u. is applied to the motor. To illustrate the current limiting capabilities of GFM-VSCs with the generalized three-phase GFM droop control during the startup of induction motor, the responses of the filter current, filter voltage, active power, reactive power, and frequency for each phase of GFM-VSC 1 and GFM-VSC 2 are depicted in Figs. 7 and 8, respectively.

Both GFM-VSCs contribute to supplying the high inrush current and reactive power demanded by the induction motor startup. However, as shown in Figs. 7 and 8, the current magnitude of each phase is effectively limited to $I_{max}$, i.e., 1.2 p.u., under unbalanced conditions. Additionally, as observed in Fig. 7, during the startup of induction motor, the frequency of the three phases remains consistently converged for GFM-VSC 1 with $k_s = 10^5$. In contrast, for GFM-VSC 2 with $k_s = 1$, the frequency of the three phases diverges during the transient period, as shown in Fig. 8.

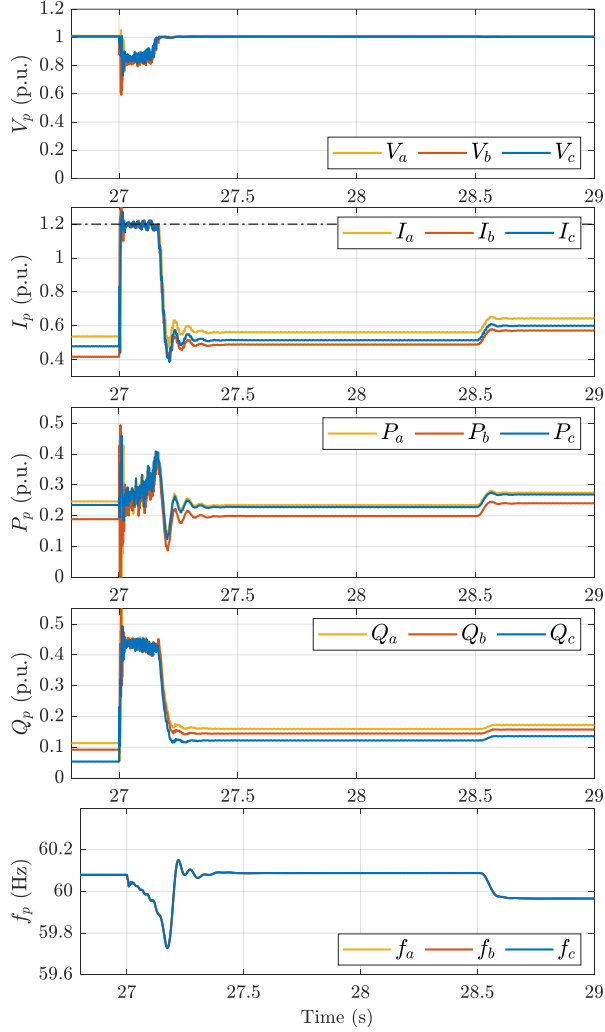

Fig. 7. The responses of the filter current, filter voltage, active power, reactive power, and frequency for each phase of GFM-VSC 1 during induction motor startup.

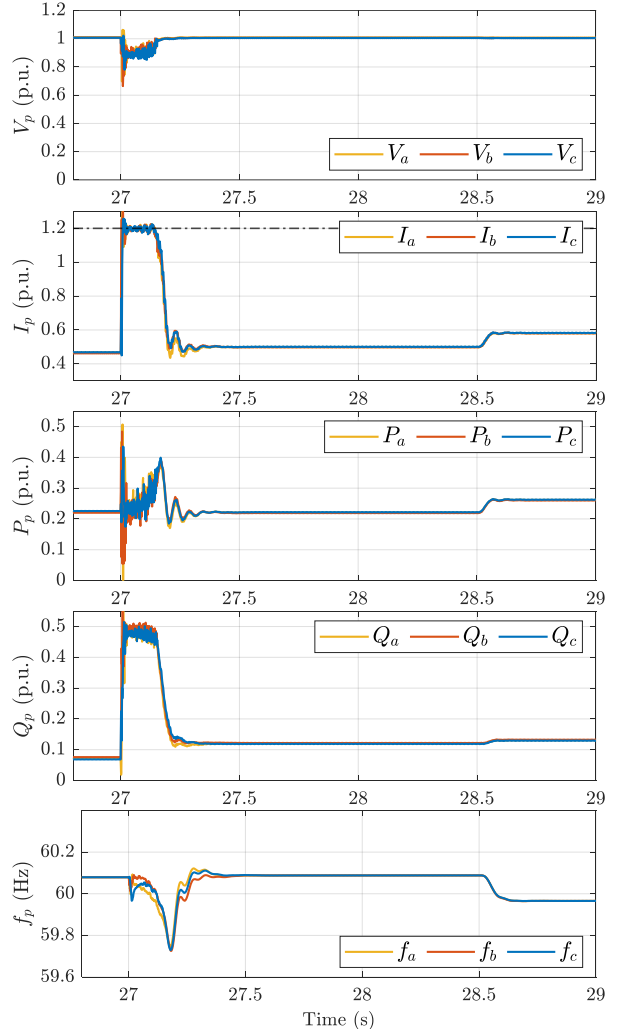

Fig. 8. The responses of the filter current, filter voltage, active power, reactive power, and frequency for each phase of GFM-VSC 2 during induction motor startup.

## 5. Conclusions

This paper investigates the black start capability of grid-forming voltage source converters (GFM-VSCs) using a generalized three-phase GFM droop control under unbalanced conditions. Through Electromagnetic Transient (EMT) simulations, the investigation focuses on the restoration of the IEEE 13-bus feeder system, incorporating advanced load relays integrated into breakers. The simulation results demonstrate the effectiveness of the generalized three-phase GFM droop control in achieving bottom-up black start operations. With this control scheme, the adjustment of phase-balancing gain $k_s$ enables a trade-off between voltage and power imbalances, ensuring accurate active power sharing among GFM-VSCs. The integration of advanced load relays into breakers facilitates a gradual reconnection of loads based on local voltage and frequency information, without the need for centralized coordination. Moreover, this paper explores the current limiting capabilities of GFM-VSCs during unbalanced fault scenarios and induction motor startups, emphasizing the effectiveness of the control in limiting current magnitudes under unbalanced conditions.

In contrast to conventional restoration schemes that rely on central planning of sequential restoration, the examined black start approach exhibits improved scalability due to the autonomous and parallel restoration capabilities of individual sectionalized feeders enabled by the presence of GFM converters. This paper serves as a proof of concept for one feeder with multiple GFM-VSCs. Conceptually, following the restoration of upstream networks, every feeder needs to synchronize with upstream and neighboring networks before reconnecting. Topics for future work include (i) assessing the scalability benefits of parallel autonomous restoration of individual feeders, and (ii) understanding the implications regarding re-synchronization with upstream and neighboring networks for full restoration.